# Distributed Data Storage in Large-Scale Sensor Networks Based on LT Codes

Saber Jafarizadeh, *Student Member, IEEE,* and Abbas Jamalipour, *Fellow, IEEE*
{saber.jafarizadeh, abbas.jamalipour}@sydney.edu.au
School of Electrical and Information Engineering, University of Sydney, Sydney NSW 2006, Australia

*Abstract*—This paper proposes an algorithm for increasing data persistency in large-scale sensor networks. In the scenario considered here, $k$ out of $n$ nodes sense the phenomenon and produced $k$ information packets. Due to usually hazardous environment and limited resources, e.g. energy, sensors in the network are vulnerable. Also due to the large size of the network, gathering information from a few central hopes is not feasible. Flooding is not a desired option either due to limited memory of each node. Therefore the best approach to increase data persistency is propagating data throughout the network by random walks.

The algorithm proposed here is based on distributed LT (Luby Transform) codes and it benefits from the low complexity of encoding and decoding of LT codes. In previous algorithms the essential global information (e.g., $n$ and $k$) are estimated based on graph statistics, which requires excessive transmissions. In our proposed algorithm, these values are obtained without additional transmissions. Also the mixing time of random walk is enhanced by proposing a new scheme for generating the probabilistic forwarding table of random walk. The proposed method uses only local information and it is scalable to any network topology. By simulations the improved performance of developed algorithm compared to previous ones has been verified.

## I. INTRODUCTION

WIRELESS sensor networks consist of a number of sensors with limited resources, e.g. energy, computational power and memory. Large number (in the order of tens of thousands or higher) of these sensors are deployed in remote and isolated environments to monitor a phenomenon e.g. bushfire or flood. In some scenarios only a few of sensors sense the phenomenon and due to the hazardous nature of environment these sensors are very vulnerable to failure. The main objective is to increase the survival chance or lifetime of data generated by these sensors. A simple solution is gathering the sensed data in one or a few central gateways. This is not possible because of network's large-scale and limited energy of each sensor. Therefore the best solution would be propagating data over network. Due to limited memory of each sensor and their vulnerability to failure, flooding or replication based algorithms are not acceptable. A much appropriate approach is storing encoded data with redundancy in sensors, similar to erasure codes. This way the original data can be recovered by gathering a certain number of encoded packets. Also due to limited memory of each node and random topology of network using deterministic routing is not possible [1].

In the model we have considered here, there are totally $n$ sensors in the network, where only $k$ of them have sensed and generated some information. Each sensor node can only store one packet. Sensors do not hold any routing tables and they don't have any knowledge about the network's topology.

Lin et al. [2] proposed the first algorithm for this scenario based on distributed fountain codes [1]. In their algorithm each sensor after sensing and generating data sends its data packet into multiple random walks. The encoding and decoding procedure is similar to the centralized LT (Luby Transform) codes [3]. The algorithm in [2] requires certain measure of global information about the network topology, including the total number of sensors ($n$), number of source nodes ($k$) and the maximum node degree (i.e., the maximum number of neighbors a node has) in network. Obtaining these information especially the latter one requires a central observation over the network. In [4] the authors have proposed a new algorithm based on LT codes which does not require maximum node degree of network. In [4] each source packet is sent on only one random walk and based on a Bernoulli trial each node XORs the received packet with the contents of its memory. In their algorithm the values of $k$ and $n$ are estimated by each node individually using the statistical properties of random walks. The estimation method proposed in [4] requires extra transmissions in addition to the actual length of random walks.

In this paper we devise an algorithm based on LT codes and random walks for distributed data storage on large-scale sensor networks. We refer to our proposed algorithm as Distributed Data Storage based on LT codes (DDSLT). The only global information that DDSLT algorithm requires is the total number of nodes ($n$) in the network, which can be easily defined for each sensor before deploying the network. The value of $k$ is obtained in each node as the random walks proceed and unlike the method in [4] for obtaining the value of $k$ no extra transmissions are required. The convergence rate of random walks to their equilibrium distribution is improved by devising a new method for generating the probabilistic forwarding table of random walks. The proposed method does not require any global knowledge and it has a better convergence rate compared to Metropolis method used in [2].

The main contributions of this paper are as follows.
1) A new algorithm is proposed for increasing data persistency in large-scale sensor networks. Our algorithm is

based on simple random walks and LT codes. It is completely local and scalable to any network topology.

2) We have calculated a lower bounds on successful encoding probability.

3) By simulations we have compare the performance of our algorithm with the previous ones. Also we show that our algorithm benefits from online decoding property of centralized LT codes, which is the main reason for using LT codes.

The organization of this paper is as follows. Section II is a brief review of related works. Preliminaries are presented in section III including the network model and a review of LT codes and random walk. In section IV, we propose DDSLT algorithm for distributed storage in large-scale sensor networks. We present the simulation results in section V and we conclude the paper in section VI.

## II. RELATED WORK

[2] is one of the first papers which employs distributed fountain codes to increase data persistency in sensor networks. In their algorithm, each node sends its data in multiple random walks. The number of random walks is determined based on the total number of nodes ($n$) and the number of sensing nodes ($k$). Each packet is saved in the cache memory of the node at the end of walk. After propagation of data is finished, each node XORes (similar to LT codes) a certain number of randomly selected data packets in its cache memory and save the resultant in its permanent memory. In [2], Metropolis method has been used for generating probabilistic forwarding table of random walk. This method requires the knowledge of maximum degree of graph, which is hard to achieve in large-scale networks. The algorithm in [2] requires huge amount of cache memory in each node for saving all received packets. Also their algorithm is not very efficient in terms of number of transmissions, since each packet is only saved at the end of random walk.

To the best of our knowledge [4] is the closest work to ours presented in this paper. Salah et al. in [4] have devised two algorithms which overcome most of disadvantages of the algorithm in [2]. In [4], each data packet is sent on one random walk and the decision on XORing of each packet is done online. This way, the extra cache memory in nodes is not required anymore and the total number of transmissions is reduced. In their algorithm the values of $n$ and $k$ are still essential for the performance of algoirthm. Therefore, they proposed a second algorithm which estimates the values of $n$ and $k$ based on the graph statistics, in particular return time of random walk. The estimation methods used in [2] requires extra transmissions in addition to the actual length of random walks. In [2] each node selects each one of its neighbors with equal probability for forwarding the packet. A disadvantage of this method is that, the stationary distribution of random walks will not match the code degree distribution of network. Therefore the final code degree distribution will not follow the desired Soliton distribution.

Dimakis et al. [1] have also proposed an algorithm for distributed data storage in sensor networks based on fountain codes. Their algorithm employs geographic routing and random walks. A major drawback of their algorithm is the requirement for geographic location of each node, which is not possible in most of sensor network applications.

Kamra et al. [5] propose Growth codes which maximizes partial data recovery at decoder when part of network becomes unavailable. Their decoding principle is similar to online decoding property of Fountain codes, but interestingly for the degree distribution of network they have obtained a distribution other than Soliton distribution.

## III. PRELIMINARIES

In this section we provide our network model for sensor network along with a brief review of LT codes [3], and random walks on graphs.

### A. Network Model

We consider a sensor network with $n$ sensors deployed randomly in the unit square $A = [0,1] \times [0,1]$. Each sensor has a transmission range of $R = 2/\sqrt{n}$ and it can send to or receive from any sensor node within its transmission range. The connectivity graph of network is a random geometric graph [6]. We assume that the connectivity graph of network is an undirected connected graph, i.e. there is a path between any two individual nodes.

We assume that only $k$ out of $n$ deployed sensors, have sensed the phenomenon. We mention these $k$ sensor nodes by source nodes and the remaining $n - k$ nodes by storage nodes. In our model we assume that source nodes are distributed uniformly among others. The proposed algorithm is independent of network topology and distribution of source nodes. Therefore its performance will be the same for the cases where the source nodes are focused in a region with a non-uniform distribution, e.g. Gaussian.

In our model each node acquires local information only from its direct neighbors. Furthermore we assume that no global information is available for nodes except the total number of nodes ($n$). This value can be loaded in nodes' memory before deploying the network.

*Definition 1: Node Degree & Maximum Node Degree*

Let $\mathcal{N}_u$ be the set of neighbors of node $u$. By neighbor we meant the nodes which can communicate directly with node $u$. The node degree of node $u$ is defined as the number of its neighbors $|\mathcal{N}_u|$ and the maximum node degree of a graph is $\max_u |\mathcal{N}_u|$.

### B. LT Codes [3]

LT codes are a member of bigger family of Fountain codes. Fountain codes are a class of rateless erasure codes. They are called rateless since unlimited number of encoded packets can be generated from a finite number of source packets. The main advantage of Fountain codes is their ability for online decoding. In other erasure codes with finite code rates, such as Reed-Solomon codes, encoding and decoding should be done in a centralized manner which is not applicable to our scenario.

Using LT codes, for generating each encoded packet, $d$ source packets are selected uniformly and XORed together. $d$ is called the code degree of encoded packet and it is determined based on a certain degree distribution. It is proved that the Ideal Soliton distribution is the optimal distribution in expectation [3]. Ideal and Robust Soliton distributions are provided in Appendix A.

In [3] Luby has shown that using LT codes, $k$ original source packets can be recovered from any $k + O(\sqrt{k}.\ln^2(k/\delta))$ encoded packets with probability $1 - \delta$. Also he has shown that the average encoding and decoding complexity is $O(\ln(k/\delta))$ and $O(k.\ln(k/\delta))$ per packet, respectively.

*C. Random Walk on a Graph*

Random walk is a random process where the next state is selected based on current state, and this selection is independent of previous states. In a random walk on a graph, nodes of a graph are the states of walk. Next state or hop of random walk is selected from the neighbors of current state or node, according to probabilistic forwarding table of random walk. A random walk can be modeled with time-reversible Markov chain, since the next state depends only on the current state. The limiting distribution of random walk stopping at a particular node is same as the steady-state distribution of Markov chain. The time (number of steps) to reach the steady state distribution is called mixing time of a Markov chain. Mixing time of Markov chain can be enhanced by adjusting its transition probability matrix. The transition probability matrix of Markov chain corresponds to probabilistic forwarding table of Random walk. The problem of enhancing Mixing time of a random walk on a graph is extensively studied by Boyd et. al. in [7].

*Definition 2: Mixing Time of Markov Chain*

Mixing time is defined as the first time (number of steps) by which the distance between the current distribution and the stationary distribution of Markov chain falls below a certain threshold.

## IV. DISTRIBUTED DATA STORAGE BASED ON LT CODES (DDSLT) ALGORITHM

In this section, we present our algorithm for disseminating sensed data among sensors in the network.

In DDSLT algorithm after sensing the phenomena, each one of source nodes, adds its ID and a counter as the header to its sensed data and sends the resultant source packet in a simple random walk. The counter is responsible for the length of random walk. This counter will be set to the desired length, by source node at the beginning of walk and it will be decreased at each transmission until it reaches zero where the walk ends. At each round, upon receiving a source packet, node $u$ will do the following procedures in turn,
  1) Updating code degree,
  2) Updating transition probabilities,
  3) XORing procedure.

In the XORing procedure based on a Bernoulli trial, node $u$ will decide on XORing the received packet with contents of its memory.

*Definition 3: Code Degree & Code Degree Distribution*

The number of source packets XORed and saved in the memory of node $u$ is called the code degree of node $u$. We refer to code degree of node $u$ as $d_u$. The distribution of $d_u$ among all nodes in the network is called code degree distribution.

After finishing all these procedures, node $u$ will adjust the value of counter responsible for the length of random walk and put the packet in its forward queue. At each transmission round, each node will transmit the first packet in its forward queue to one of its neighbors according to its transition probabilities.

To achieve the desired code degree distribution, it is crucial for each source packet to visit each node in the network at least once. Thus we choose the length of random walk same as the cover time of the network's connectivity graph. The cover time for a random walk on a graph is defined as follows.

*Definition 4: Cover Time of a Random Walk*

Let $C_v$ denote the expected number of steps, that every node has been visited by the walk, starting from node $v$. Then the cover time of the graph is defined as

$$T_c(G) = \max_{v \in G} C_v.$$

*Lemma 1:* [8]

With high probability, the cover time of a connected random graph is bounded by

$$T_c(G) = \Theta(n.\ln(n)).$$

According to Lemma 1, the value of counter responsible for length of random walk is set to $C_1.n.\ln(n)$ in the source node, at the beginning of random walk. At each round this counter decreases by one until it reaches zero, where the walk ends. The counter assures that each node receives the packet at least once. $C_1$ is a system coefficient to be determined.

In the following we explain, in detail, three procedures mentioned above, namely, updating code degree, updating transition probabilities and XORing procedure.

*A. Updating Code Degree Procedure*

In DDSLT algorithm each node $u$ has a counter $k'_u$ for the number of different source node IDs that it receives. Node $u$ determines its code degree based on $k'_u$ and $\alpha_u$, where $\alpha_u$ is a random number selected uniformly in the interval $[0,1]$. $\alpha_u$ is saved during the time that node $u$ is functioning. We define $\Omega(k)$ as the Cumulative Distribution Function (CDF) of the desired degree distribution. Since $k$ is not available to any of the nodes, then node $u$ forms the $\Omega(k)$ using $k'_u$ as an estimation of $k$. Then node $u$ checks the intervals in $\Omega(k)$ and selects the value of its code degree ($d_u$) based on the interval containing $\alpha_u$ within. For example let $k = 3$ then the intervals corresponding to $d_u = 1, 2, 3$ would be $[0, 1/3]$, $[1/3, 5/6]$, $[5/6, 1]$. If $\alpha_u = 0.8147$ then $d_u$ would be selected equal to 2.

*Lemma 2:*

Let $\Omega(k)$ be one of either Ideal $(\Omega_{is}(k))$ or Robust $(\Omega_{rs}(k))$ Soliton distributions. Then using the updating code degree procedure, mentioned above, the value of $d_u$ increases monotonically, as the random walk continues.

*Proof.*

At the start of random walk, the value of $k'_u$ is zero. Assuming that all packets visit each node at least once then the final value of $k'_u$ would be equal to $k$. Without loss of generality, we assume that $\Omega(k)$ is the CDF of Ideal Soliton distribution and $\alpha_u$ is in the interval corresponding to degree $d_u$. Increasing $k'_u$ by one, the bounds of all intervals will decrease. Since $\alpha_u$ is a fixed number then it will be within one of the new intervals corresponding to either $d_u$ or $d_u + 1$, and thereafter Lemma 2 holds. The proof for the case of Robust Soliton distribution can be followed similarly.

$\alpha_u$ has a uniform distribution among all nodes in the network, then if $k'_u$ in all nodes reaches $k$, the code degree distribution would be same as the desired one. This necessitates that all random walks should visit all nodes in the network at least once, which might not feasible in most cases. As a partial solution to this problem we add a source packet counter $(SC_j)$ to the header of each source packet $j$. When node $u$ receives the source packet $j$, it checks $SC_j$, if it exceeds the number of source IDs identified in node $u$ then node $u$ will modify $k'_u$ to $SC_j$, otherwise it will increase $k'_u$ by one and set $SC_j$ equal to the new value of $k'_u$. This way the value of $k'_u$ in nodes will reach the value of $k$ faster. The number of source IDs recognized in node $u$ is identified by $SN_u$.

By following simulations, we investigate the performance of proposed updating code degree procedure. In Fig. 1 the percentage of nodes which their source counters $(k'_u)$ have reached the value of $k$ is depicted in terms of $C_1$. $C_1$ is the system parameter related to length of each random walk $(C_1.n.\ln(n))$. The network parameters considered for this simulation are $n = 100$, $k = 10$. We have considered same network with different communication radius $(r/\sqrt{n})$ for each node to show the relation between the performance of updating procedure with the density of graph. Higher values of $r$ will result in more edges in graph.

It is obvious from Fig. 1 that almost all of the nodes reach the actual value of $k$ before $C_1 = 1$. In our simulations and also in [4] $C_1$ has been chosen equal to 5. Therefore nodes do not have to wait until the end of random walk to reach the exact value of $k$. Also nodes reach the value of $k$ faster when their mean node degree is higher.

### B. Updating Transition Probabilities

Each one of random walks can be modeled as a Markov chain. The transition probability of random walk from node $u$ to node $v$ is the element $(u, v)$ of transition probability matrix $(TP)$ of Markov chain. For generating the transition probability matrix of Markov chain we use the following formulas.

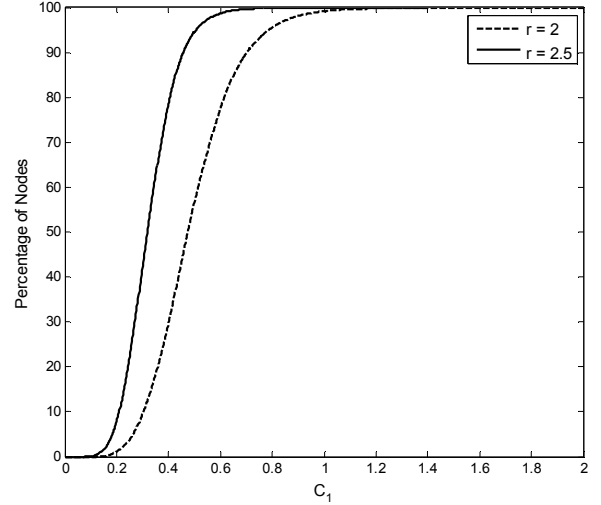

Fig. 1. Percentage of nodes which are reaching the value of $k$ in terms of $C_1$ (Length of random walk).

$$TP_{u,v} = \min\left(\mu_v, \mu_u \times \frac{d_v}{d_u}\right) \text{ for } u \neq v,$$

$$TP_{u,u} = 1 - \sum_{\substack{v \in \mathcal{N}(u) \\ u \neq v}} P_{u,v} \quad (1)$$

with $\mu_u$ defined as

$$\mu_u = \frac{d_u}{\sum_{\ell \in \mathcal{N}(u)} d_\ell}$$

The only information required for generating transition probabilities (1) is the code degree of neighboring nodes. It is obvious that this method for generating the transition probabilities does not require any global information and it is scalable to any network.

As explained in section IV-A, the value of code degree $d_u$ in each node gets updated by time. Since the transition probabilities depend on the code degree of nodes, they will be affected by the updating procedure of code degrees. This effect has been explained by following lemma.

*Lemma 3:*

Let $d_u$ be the code degree of node $u$. The effect of any changes in $d_u$ is limited to the transition probabilities of edges up to the second order neighborhood of node $u$.

*Proof.*

Changes in $d_u$ will change only $\mu_u$ and $\mu_v: \forall v \in \mathcal{N}(u)$. Therefore the only transition probabilities that may be affected are the transition probabilities of edges connected to node $u$ and its neighboring nodes.

In the following we have compared the methods proposed for designing transition probabilities in [2, 4], with the one proposed here (1).

The proposed method in [4] for determining the transition probabilities is also local but the stationary distribution of transition probability matrix is not based on code degree of nodes. In their method each node selects each one of its neighbor with equal probability. Thus the transition probability for the edges connected to node $u$ is $P_{u,v} = 1/|\mathcal{N}(u)|$. Also the stationary distribution for node $u$ would be $|\mathcal{N}(u)|$, which is not the one selected based on desired code degree distribution. Also their method requires the global knowledge regarding $k$ in the next part of algorithm for selecting source packets. The authors in [2] have used the following formula for designing the transition probabilities, which is based on Metropolis algorithm.

$$TP_{u,v} = \frac{\min(1, d_v/d_u)}{D_{max}} \quad \text{for} \quad u \neq v \quad (2)$$

where $D_{max} = \max_v |\mathcal{N}(v)|$ is the maximum degree of graph. $D_{max}$ is a global knowledge and it is hard to achieve in large-scale networks.

Disregard of these methods' requirement for global knowledge, a major difference between these three methods is their convergence rate to the stationary distribution. The higher convergence rate results in faster (less) mixing time, thus shorter length for random walks and less transmissions. The convergence rate of a Markov chain has a reverse relation with its mixing time.

*Lemma 4:*[7]

The asymptotic convergence rate of Markov chain to its stationary distribution is determined by the Second Largest Eigenvalue Modulus (SLEM) of its transition probability matrix ($TP$). The smaller SLEM results in faster convergence.

The SLEM of the transition probability matrix ($TP$) is defined as follows.

*Definition 5: SLEM* [7]

Let $-1 \leq \lambda_n \leq \cdots \leq \lambda_2 < \lambda_1 = 1$, be the eigenvalues of transition probability matrix ($TP$) in non-increasing order. Then the SLEM of $TP$ is defined as following.

$$SLEM(TP) = \max\{\lambda_2, -\lambda_n\}$$

In Table 1 we have compared two methods proposed in [2, 4] for designing transition probabilities with (1). Our comparison is based on the SLEM value of obtained transition probability matrices. All three methods are compared over random geometric graphs with 100 nodes. Also the stationary distribution of both methods (1) and (2) are the same and it is derived from Ideal Soliton distribution.

Table 1. SLEM of transition probability matrices obtained from (1), (2) and [4].

| Method | (2) | [4] | (1) |
|---|---|---|---|
| SLEM | 0.9900 | 0.9689 | 0.9788 |

From Table 1 one can conclude that the SLEM of method proposed in [4] is smaller and its convergence is faster. But the stationary distribution is not the same as code degree distribution of nodes, which downgrades the overall performance of algorithm.

### C. XORing Procedure

Here we explain, how in DDSLT algorithm, each node will accept the received packet and XOR it with the contents of its memory. The main goal is to increase the probability that the number of XORed source packets at each node will be exactly equal to the node's code degree.

When a storage node $u$ receives its first packet, it will save the packet to its memory with probability one. After receiving the second packet, node $u$ will form $k'_u$ and $d_u$ as described in section IV-A. Then it will run the Bernoulli process for the first packet and with probability $d_u/k'_u$ it will save the packet. If the code degree $d_u$ of node is not fulfilled then it will run the Bernoulli process with the same probability for the second packet and XOR the outcome with contents of memory.

For the rest of the packets (after second received packet) the procedure for all nodes is as following.

Upon receiving the source packet $j$ at node $u$, after updating $k'_u$ and $d_u$, the node will check the number of source packets XORed and saved in its memory. If $d_u$ exceeds this number and packet $j$ has not been XORed previously. Then node $u$ will run a Bernoulli process and with probability $d_u/k'_u$ it will XOR packet $j$ with the contents of its memory. Following this procedure and considering the fact that $d_u$ is a monotonically increasing function of time (Lemma 2), we can conclude that the number of XORed packets in a node will never exceed its code degree $d_u$. In [2, 4] nodes only enter a packet in Bernoulli process if it is the packet's first visit. This way there is a probability that a node might not be able to XOR enough number of packets to fulfill its code degree. In our method this probability is much less since each packet enters the Bernoulli process in its every visit.

We assume that each source node saves its original source packet which is considered as the first packet. Therefore source nodes treat all packets in the same way.

### D. DDSLT in Detail

In the following we describe the Initialization, Dissemination and Encoding phases of DDSLT algorithm in detail. In Table 2 we have listed the variables, used in DDSLT algorithm.

Table 2. List of Variables.

| Variable | Description |
|---|---|
| $k'_u$ | Estimated number of source packets in node $u$ |
| $d_u$ | Code degree of node $u$ |
| $Sd_u$ | Number of source packets XORed with the contents of memory of node $u$. |
| $SN_u$ | Number of different source IDs identified at node $u$ |
| $SC_j$ | Source Counter of packet $j$. The counter on the header of packet $j$ which saves the estimation of $k$. |

*Initialization phase:*

- Each node $u$ selects a random number ($\alpha_u$) according to uniform distribution in the interval [0,1] as explained in section IV-A. We assume that the functions for generating the CDF of desired degree distribution, (i.e. Ideal or Robust Soliton distributions (Appendix A)), are predefined for each node.

- After sensing, each source node $s_i$, $i = 1, ..., k$ creates its source packet as following.

$$P_{s_i} = [ID_{s_i}, L_{s_i}, SC_{s_i}, UF_{s_i}, x_{s_i}],$$

$ID_{s_i}$ is the ID of Source node $s_i$. $L_{s_i}$ is the counter determining the length of random walk which is set to $C_1.n.\ln(n)$. $SC_{s_i}$ is the source counter of each packet as explained in section IV-A. $SC_{s_i}$ is set to one. $UF_{s_i}$ is the update flag determining if the packet is a new packet or an update for a previously sent packet and $x_{s_i}$ is the actual information generated by the source node $s_i$. After forming the source packet, each source node forwards the packet to one of its neighbors $\mathcal{N}(s_i)$ according to its transition probabilities.

- Each one of storage and source nodes sets their default value of variables as following. By default we assume that each source node will XOR and save its source packet.

| Storage Node $u$ |
|---|
| $k'_u = 0$ |
| $d_u = 1$ |
| $Sd_u = 0$ |
| $SN_u = 0$ |

| Source Node $u$ |
|---|
| $k'_u = 1$ |
| $d_u = 1$ |
| $Sd_u = 1$ |
| $SN_u = 1$ |

*Dissemination and Encoding phase:*

For storage node $u$, dissemination and encoding phase is as follows.

- Upon receiving the first packet $(P_{S_i})$, node $u$ will save the information content of packet $(x_{s_i})$ to its memory with probability one. Then it will
  - increase $SN_u$ to one,
  - update the value of $k'_u$ to the value of source counter of packet $(SC_i)$,
  - decrease the value of $L_{s_i}$ by one and put the packet in its forward queue.

- After receiving the second new packet $(P_{s_j})$ for the first time, node $u$ will
  - increase $SN_u$ to two,
  - update the value of $k'_u$ and the source counter of packet $(SC_j)$ to max $(k'_u, SN_u, SC_j)$,
  - based on the value of $\alpha_u$ and $k'_u$, node $u$ will form its code degree $(d_u)$,
  - start the Bernoulli process for the first packet and saves the information content of packet $(x_{s_i})$ with probability $(d_u/k'_u)$ and depending on the outcome of Bernoulli process, node $u$ will increase $Sd_u$ from zero to one.

- if $Sd_u < d_u$, it will run the Bernoulli process with the same probability of success for the second packet, and depending on the outcome, it will XOR $x_{s_j}$ with the contents of its memory and it will increase $Sd_u$ by one,
- decrease the value of $L_{s_j}$ by one and put the packet in its forward queue.

- For the rest of packets, after receiving the packet $(P_{s_k})$, node $u$ will
  - check the source node ID $(ID_{s_k})$ of the packet. If it is the first visit of packet, it will increase $SN_u$ by one,
  - update the value of $k'_u$, $SC_{s_k}$ and $d_u$,
  - if the code degree is not fulfilled ($Sd_u < d_u$), and packet has not been saved before, then run the Bernoulli process and with probability $(d_u/k'_u)$ XOR the packet $(x_{s_k})$ with the contents of its memory, and it will increase $Sd_u$ by one, if the packet is XORed.
  - decrease the value of $L_{s_k}$ by one and put the packet in its forward queue.

For source nodes, we assume that each source node will save its own data by default. Therefore after receiving the second source packet, each source node will continue the same procedure as the storage nodes.

*E. Final Code Degree Distribution*

Here we investigate the probability that each node fulfils its code degree. This is equivalent to the probability of successful encoding.

We assume that packet $j$ visits node $u$, $N_j(u)$ times during the walk. The probability that packet $j$ would not be accepted in node $u$, $\left(P_j^R(u)\right)$ during the whole walk is

$$P_j^R(u) = \prod_{i=1}^{N_j(u)} \left(1 - \frac{d_u(t_i)}{k'_u(t_i)} \times sgn(d_u(t_i) - Sd_u(t_i))\right),$$

where $t_i$ is the time at $i$-th visit of packet $j$ to node $u$ and function $sgn(x)$ is defined as follows.

$$sgn(x) = \begin{cases} 1 & \text{for } x > 0, \\ 0 & \text{for } x \leq 0. \end{cases}$$

The function $sgn(d_u(t_i) - Sd_u(t_i))$ is for the constraint on nodes not to save additional packets if their degree is fulfilled. Relaxing this constraint the probability that packet $j$ would not be accepted in node $u$, will reduce to

$$P_j^R(u) = \prod_{i=1}^{N_j(u)} \left(1 - \frac{d_u(t_i)}{k'_u(t_i)}\right).$$

Since $\forall t_i : \frac{d_u(t_i)}{k'_u(t_i)} \geq \frac{d_u}{k}$ then

$$P_j^R(u) \leq \left(1 - \frac{d_u}{k}\right)^{N_j(u)}.$$

Let $P_j^A(u)$ be the probability that packet $j$ will be accepted in node $u$. Then for $P_j^A(u)$ we have

$$P_j^A(u) = 1 - P_j^R(u) \geq 1 - \left(1 - \frac{d_u}{k}\right)^{N_j(u)}.$$

Since packets are running in independent random walks, then the probability that $d_u$ packets will be saved in node $u$ at the end of DDSLT algorithm is as following.

$$Pr(Sd_u = d_u) = \Omega(d_u) \times \prod_{i=1}^{d_u} P_{j_i}^A(u)$$

$$\geq \Omega(d_u) \times \prod_{i=1}^{d_u} \left(1 - \left(1 - \frac{d_u}{k}\right)^{N_{j_i}(u)}\right)$$

where $j_1, \ldots, j_{d_u}$ are the accepted packets. It should be mentioned that there are $\binom{k}{d_u}$ different combinations for selecting the $d_u$ packets out of possible $k$ packets.

Let $L = C_1.n.\ln(n)$ be the length of each random walk. Neglecting the effect of transient time of random walk, after the random walk has reached its stationary distribution, the expected number of visits of packet $j$ to node $u$ is $E[N_j(u)] = L \times \frac{d_u}{\Sigma_d}$, where $\Sigma_d = \sum_v d_v$. Replacing $N_{j_i}(u)$ by its expected value, for $Pr(Sd_u = d_u)$, we have:

$$Pr(Sd_u = d_u) \geq \Omega(d_u) \times \binom{k}{d_u} \times \left(1 - \left(1 - \frac{d_u}{k}\right)^{L \times \frac{d_u}{\Sigma_d}}\right)^{d_u}$$

$Pr(Sd_u = d_u)$ is the probability that node $u$ will successfully fulfil its code degree. But the above bound has been achieved by relaxing the constraint that, nodes cannot save new packets if their code degree is fulfilled. This constraint does not have any effect on the probability that node $u$ saves less than $d_u$ packets. Therefore this constraint simply increases the probability that node $u$ will save $d_u$ packets and increases the probability of successful encoding. Also it can be concluded that the above bound is not very tight.

The total number of transmissions is upper bounded by $k.C_1.n.\ln(n)$, since a number of diagonal elements in transition probability matrix are positive. Therefore there is a chance that a packet might stay in the same node for next transmission time.

In [3] Luby proves that for LT codes if Robust Soliton distribution is used as the code degree distribution, $k$ original source blocks can be recovered from any $k + O(\sqrt{k}.\ln^2(k/\delta))$ encoded output blocks with probability $1 - \delta$. Also he has shown that the complexity of both encoding and decoding is $O(k.\ln(k/\delta))$. Since we have used Soliton distribution as the desired code degree distribution, we expect the same decoding capability and encoding/decoding complexity of LT codes.

### F. Updating Data

In this section we consider the case when encoding is done and the source node $s_i$ wants to update its data among all nodes who has stored its data. Our algorithm can employ the same updating procedure introduced in [4]. The source node $s_i$ prepares a packet as following, sets the update flag $(UF_{s_i})$ and sends the packet in a random walk same as the previous ones.

$$P_{s_i} = [ID_{s_i}, L_{s_i}, SC_{s_i}, UF_{s_i}, x_{s_i} \otimes x'_{s_i}].$$

$x_{s_i}$ and $x'_{s_i}$ are the old and new data, respectively. It should be mentioned that this updating scheme requires an extra memory in source nodes to store its source packet in addition to its encoded packet. This extra memory might not be possible in some scenarios. A solution to this problem would be removing the encoded packet from the source nodes. This means that each source node will simply save only its own source packet. This approach will affect the decoding capability of algorithm, since the number of nodes with code degree one will increase which in turn will widen the gap between actual code degree distribution and Soliton distribution.

### V. SIMULATIONS

In this section we investigate the performance of DDSLT algorithm by simulations. To have a fare comparison with the previous developed algorithms in [4], we select the performance metrics same as those used in [4]. We plot the successful decoding probability in terms of decoding ratio. The decoding ratio $(\eta)$ is defined as the number of encoded packets collected in decoder $(h)$ divided by the number of source packets $(k)$.

$$\eta = \frac{h}{k}.$$

Successful decoding probability is defined as the probability that all $k$ source packets can be recovered from the $h$ collected packets. Let $M = \binom{n}{h}$ be the number of all possible combinations for selecting $h$ packets from $n$ possible packets, and $M_s$ be the number of combinations that $k$ source packets can be recovered. Then the successful decoding probability would be $M_s/M$. Of course testing all $M$ combinations are not possible therefore only a random portion of them are tested for the results presented in this section.

In Fig. 2 we have compared DDSLT algorithm with LTCDS-I algorithm proposed in [4]. The network considered for this simulation has $n = 100$ nodes with $k = 10$ source nodes. The length of each random walk is $5 \times n.\ln(n)$. It should be mentioned that in DDSLT algorithm only the value of $n$ is required at each node while in LTCDS-I, both values of $n$ and $k$ should be known at each node.

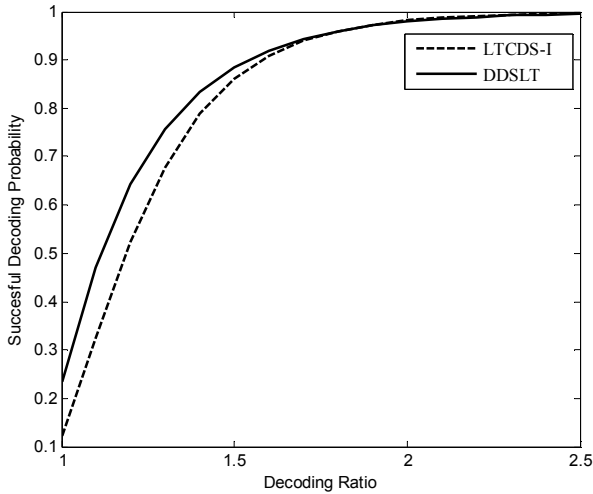

Fig. 2. Successful decoding probability in terms of decoding ratio ($\eta$) for DDSLT algorithm and LTCDS-I proposed in [4].

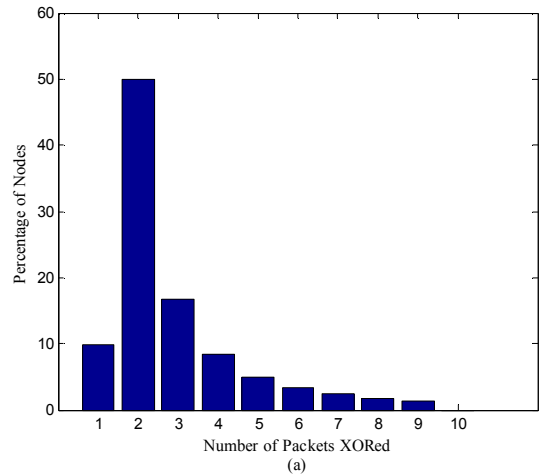

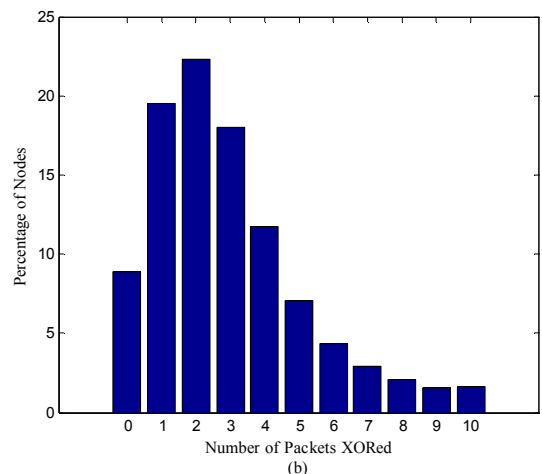

Fig. 3. Final distribution of number of XORed packets in nodes using (a) DDSLT algorithm and (b) LTCDS-I algorithm, over a network with $n = 100$, $k = 10$, $C_1 = 5$.

From Fig. 2 it is obvious that DDSLT algorithm has a better performance compared to LTCDS-I, especially for decoding ratios less than 2. But for high decoding ratios both algorithms have almost the same performance. The reason is the method used for measuring the successful decoding probability. Successful decoding probability is obtained by checking the rank of matrix of coefficients (with elements 0 and 1) for the encoded packets collected in decoder. If the rank reaches $k$ then all original source packets can be decoded and the decoding is considered as successful. In order to analyze this matter in detail in Fig. 3 we have compared the distribution of number of actual XORed packets in nodes.

It is obvious from Fig. 3 that the final distribution using DDSLT algorithm is much closer to Ideal Soliton distribution compared to the distribution obtained from LTCDS-I. Using LTCDS-I algorithm almost 10 percent of nodes do not XOR any of source packets while the number of nodes which have XORed up to 8, 9 or 10 packets are much higher than the value initiated by Ideal Soliton distribution. The high number of packets with higher degree increases the rank of matrix of coefficients for the encoded packets collected in decoder. This means that the decoding procedure of packets obtained from LTCDS-I cannot be same as LT codes and it will require matrix inversion and extra computation.

In conventional LT codes there is a high probability that a number of source packets will be decoded before realization of all encoded packets in decoder. This property is called online decoding property of LT codes and it reduces the computational complexity of decoding procedure. The probability of successful online decoding depends on the number of encoded packets with low code degree. Ideal Soliton distribution is the optimal distribution, in expectation, for enhancing the online decoding property of LT codes [3]. It is obvious that LTCDS-I algorithm does not benefit from this property of LT codes due to its unbalanced distribution among the received encoded packets. As mentioned in section IV-B the stationary distribution of transition probability matrix used in LTCDS-I is not same as the degree distribution of nodes initiated by Soliton distribution. This is one of the reasons for the difference between the actual and the desired degree distribution of encoded packets in LTCDS-I algorithm.

Two main reasons that DDSLT algorithm is reaching the Ideal Soliton distribution are as following. I) each packet enters the Bernoulli process in its every visit to each node unless it is already XORed by the contents of node. This increases the probability that each node will fulfill its code degree. II) Each node does not XOR more than its code degree which upper limits the number of XORed packets in each node. It is obvious that DDSLT algorithm benefits from the online decoding property of LT codes and its low decoding computation.

In DDSLT algorithm the only global information required for each node is $n$, which is only used to set the length of random walks. To analyze the impact of correct value of $n$ on the performance of DDSLT algorithm, in Fig. 4 the percentage of nodes fulfilling their code degree versus time (Number of steps) is depicted. The percentage of nodes fulfilling their code degree is a measure of how much of the encoding process has been done.

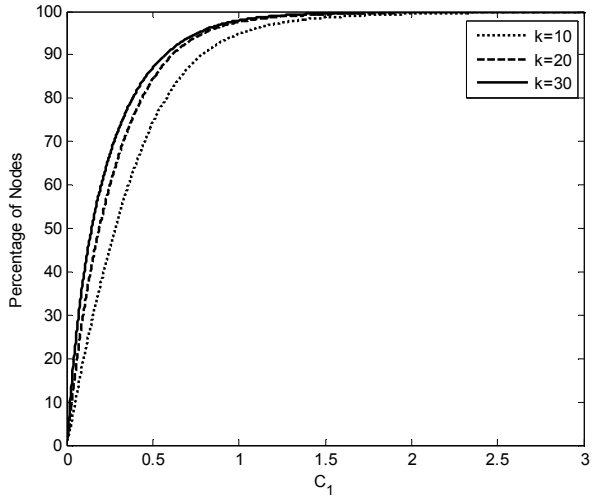

Fig. 4. The percentage of nodes fulfilling their code degree, using DDSLT algorithm, versus the number of steps (time) for $n = 100$.

It is obvious from Fig. 4 that encoding is almost finished until step $2.5 \times n.\ln(n)$. This means that one can set the value of $C_1$ to 2.5 to have shorter random walks and less transmissions. But there is a tradeoff between the value of $C_1$ and the sensitivity of algorithm to each node's estimation from $n$. In the cases which a source node's estimation of $n$ is less than its actual value, the probability that the corresponding source packet will not visit all nodes increases. This results in lower probability of successful encoding. Thus the sensitivity of algorithm to the value of $n$ decreases at the cost of longer random walks and more transmissions.

## VI. CONCLUSION

In this paper, we consider wireless sensor networks consisting of, large number of sensors with limited memory and computational power, which are vulnerable to failure. We have devised an algorithm for increasing data persistency in large-scale WSN by distributing the sensed data throughout the network. Our algorithm is completely robust and distributed. The most important advantage of our algorithm is its independency from network topology and scalability to any network. Previously developed algorithms required global information about the network topology such as maximum node degree of network, number of data packets $(k)$ and the total number of nodes $(n)$ [2, 4]. While in our algorithm the only global information required is the total number of nodes $(n)$. By simulations we confirm the superior performance of our algorithm compared to others and we show that our algorithm benefit from online decoding property of LT codes. Also we have investigated the tradeoff between the length of random walks and nodes' estimations from $n$.

## APPENDIX A
### IDEAL & ROBUST SOLITON DISTRIBUTIONS [3]

Ideal Soliton Distribution: Let $K$ be the number of data packets, then the ideal Soliton distribution is defined as following.

$$Pr(d = i) = \rho(i) = \begin{cases} \dfrac{1}{K} & \text{for } i = 1 \\ \dfrac{1}{i(i-1)} & \text{for } i = 2, \dots, K \end{cases}$$

Robust Soliton Distribution: Let $R = c.\ln(K/\delta).\sqrt{K}$ for some constant $c > 0$ and

$$\tau(i) = \begin{cases} R/i.k & \text{for } i = 1, \dots, k/R - 1, \\ R.\dfrac{\ln(R/\delta)}{k} & \text{for } i = k/R, \\ 0 & \text{for } i = k/R, \dots, k, \end{cases}$$

then the Robust Soliton distribution is defined as

$$Pr(d = i) = \mu(i) = \dfrac{\rho(i) + \tau(i)}{\beta},$$

where $\beta = \sum_i \rho(i) + \tau(i)$.